\documentstyle[12pt,epsf]{article}

\input{epsf}

\def\spose#1{\hbox to 0pt{#1\hss}}
\def\lsim{\mathrel{\spose{\lower 3pt\hbox{$\mathchar"218$}}
 \raise 2.0pt\hbox{$\mathchar"13C$}}}
\def\gsim{\mathrel{\spose{\lower 3pt\hbox{$\mathchar"218$}}
 \raise 2.0pt\hbox{$\mathchar"13E$}}}

\begin{document}

\begin{titlepage}

\begin{flushright}
CERN-TH/97-307\\
TUM-HEP-300/97\\
TTP97-41\\
hep-ph/9711262
\end{flushright}
\vspace{0.2truecm}
\begin{center}
\boldmath
\large\bf
Penguin Topologies, Rescattering Effects and\\ 
\vspace{0.2truecm} 
Penguin Hunting with $B_{u,d}\to K\overline{K}$ and $B^\pm\to\pi^\pm K$ 
\unboldmath
\end{center}
\vspace{0.1cm}
\begin{center}
Andrzej J. Buras\\[0.1cm]
{\sl Technische Universit\"at M\"unchen, Physik Department\\ 
D-85748 Garching, Germany}\\[0.6cm]
Robert Fleischer\\[0.1cm]
{\sl Theory Division, CERN, CH-1211 Geneva 23, Switzerland}\\[0.6cm]
Thomas Mannel\\[0.1cm]
{\sl Institut f\"ur Theoretische Teilchenphysik, Universit\"at Karlsruhe\\
D-76128 Karlsruhe, Germany}
\end{center}
\vspace{0.1cm}
\begin{abstract}
\vspace{0.2cm}\noindent
In the recent literature, constraints on the CKM angle $\gamma$ arising from 
the branching ratios for $B^\pm\to\pi^\pm K$ and $B_d\to\pi^\mp K^\pm$ 
decays received a lot of attention. An important theoretical limitation 
of the accuracy of these bounds is due to rescattering effects, such as 
$B^+\to\{\pi^0K^+\}\to\pi^+K^0$. We point out that these processes are 
related to penguin topologies with internal up quark exchanges and derive
$SU(2)$ isospin relations among the $B^+\to\pi^+K^0$ and $B_d^0\to\pi^-K^+$ 
decay amplitudes by defining ``tree'' and ``penguin'' amplitudes in a 
proper way, allowing the derivation of generalized bounds on the CKM
angle $\gamma$. We propose strategies to obtain insights into the dynamics 
of penguin processes with the help of the decays $B_{u,d}\to K\overline{K}$ 
and $B^\pm\to\pi^\pm K$, derive a relation among the direct CP-violating 
asymmetries arising in these modes, and emphasize that rescattering effects 
can be included in the generalized bounds on $\gamma$ completely this way. 
Moreover, we have a brief look at the impact of new physics.
\end{abstract}

\vfill
\noindent
CERN-TH/97-307\\
Revised version\\
May 1998

\end{titlepage}

\thispagestyle{empty}
\vbox{}
\newpage
 
\setcounter{page}{1}

\section{Introduction}\label{intro}
As was pointed out in \cite{PAPIII}--\cite{groro}, the decays 
$B^+\to\pi^+K^0$, $B_d^0\to\pi^- K^+$ and their charge-conjugates may play 
an important role to determine the angle $\gamma$ of the usual 
non-squashed unitarity triangle \cite{ut} of the Cabibbo--Kobayashi--Maskawa 
matrix (CKM matrix) \cite{ckm} at future $B$-factories (BaBar, BELLE, 
CLEO III; interesting feasibility studies can be found in 
\cite{groro,wuegai,babar}). The corresponding decay amplitudes can be 
expressed as 
\begin{eqnarray}
A(B^+ \to \pi^+K^0) &=& P^{(s)} + 
c_d\, P_{\rm EW}^{(s){\rm C}} + A^{(s)}\label{ampl-char} \\
A(B^0_d \to\pi^- K^+) &=& - \left[\left(\, {\cal P}^{(s)} + 
c_u\, {\cal P}_{\rm EW}^{(s){\rm C}}\,\right) + 
{\cal T}^{(s)}\right],\label{ampl-neut}
\end{eqnarray}
where $P^{(s)}$, ${\cal P}^{(s)}$  denote QCD penguin amplitudes, 
$P_{{\rm EW}}^{(s){\rm C}}$, ${\cal P}_{{\rm EW}}^{(s){\rm C}}$
correspond to ``colour-suppressed'' electroweak penguin contributions, 
$A^{(s)}$ is due to annihilation processes, and ${\cal T}^{(s)}$ is usually 
referred to as a ``colour-allowed'' $\bar b\to\bar uu\bar s$ ``tree'' 
amplitude. The label $s$ reminds us that we are dealing with 
$\bar b\to\bar s$ modes, the minus sign in (\ref{ampl-neut}) is due to our 
definition of meson states, and $c_u=+2/3$ and $c_d=-1/3$ are the up- and 
down-type quark charges, respectively.  

The CLEO collaboration has recently reported the first results for the 
combined, i.e.\ averaged over decay and charge-conjugate, branching 
ratios $\mbox{BR}(B^\pm\to\pi^\pm K)$ and $\mbox{BR}(B_d\to\pi^\mp K^\pm)$
\cite{cleo}. These quantities may lead to interesting constraints on the CKM 
angle $\gamma$, if their ratio 
\begin{equation}\label{Def-R}
R\equiv\frac{\mbox{BR}(B_d\to\pi^\mp K^\pm)}{\mbox{BR}(B^\pm\to\pi^\pm K)}
\end{equation}
is found experimentally to be smaller than 1 \cite{fm2}. Since the present
CLEO data give $R=0.65\pm0.40$, this may indeed be the case. The bounds on 
$\gamma$ obtained in this manner turn out to be complementary to the present 
range for this angle arising from the usual fits of the unitarity 
triangle (for a review, see for instance \cite{bf-rev}), and are hence of 
particular phenomenological interest. A detailed analysis of the implications 
of these bounds for the determination of the unitarity triangle has been 
performed in \cite{gnps}. 

An important limitation of the theoretical accuracy of the ``na\"\i ve''
bounds on $\gamma$ derived in \cite{fm2} -- besides the ``colour-suppressed'' 
electroweak penguin contributions -- is due to rescattering processes of the 
kind $B^+\to\{\pi^0K^+,\,\pi^0K^{\ast +},\,\rho^0K^{\ast +},\,\ldots\,\}\to
\pi^+K^0$, which have received a lot of attention in the recent literature 
\cite{gewe}--\cite{rf-FSI} (for earlier references, see \cite{FSI}). In
this paper, we focus on these rescattering effects. Following closely 
\cite{PAP0,rev}, we show in Section~\ref{pen-zoo} that they are related 
to penguin topologies with internal up quark exchanges. Analogously, 
rescattering processes such as $B^+\to\{\overline{D^0}D_s^+\}\to\pi^+K^0$ 
can be regarded as long-distance contributions to penguins with internal 
charm quarks, which also received considerable interest in the recent 
literature~\cite{PAP0}--\cite{lno}. 

Our paper is organized as follows: in Section 2, we discuss the penguin
topologies in general terms. In particular, we establish the relation 
between the penguin diagram pictures used by CP practitioners and the 
formal operator method. Here we also recall useful expressions for the 
$\bar b\to\bar s$ and $\bar b\to\bar d$ penguin amplitudes. In 
Section~\ref{pen-zoo}, we derive a simple isospin relation between
the $B^\pm\to\pi^\pm K$ and $B_d\to\pi^\mp K^\pm$ decay amplitudes by
defining the ``tree'' and ``penguin'' amplitudes in a proper way, and
address the question of rescattering effects in penguin-induced $B$ decays.
These isospin relations play a key role to probe $\gamma$ and allow the 
derivation of generalized bounds on this CKM angle. In Section~\ref{pen-hunt},
we point out that decays of the type $B_{u,d}\to K\overline{K}$ play an 
important role to obtain insights into rescattering processes and the 
dynamics of penguin topologies. Using their combined branching ratios and 
BR$(B^\pm\to\pi^\pm K)$, interesting relations and bounds can be derived, 
including a relation between the direct CP-violating asymmetries arising in 
these modes. Moreover, it is even possible to take into account the 
rescattering effects {\it completely} in the generalized bounds on $\gamma$ 
by following these lines. After a brief look at the impact of new-physics 
contributions to $B^0_d$--$\overline{B^0_d}$ mixing in Section~\ref{new-phys},
we collect our conclusions in Section~\ref{concl}. Technical details related 
to the isospin structure of the relevant amplitudes are relegated to an 
appendix.

\section{Penguin Topologies}\label{pen-topo}
In many phenomenological analyses of $B$ decays in the literature,
it is customary to represent penguin contributions by penguin diagrams
with explicit $W^{\pm}$, $t$, $c$ and $u$ exchanges, as shown in Fig.\ 1 (a). 
On the other hand, the proper treatment of $B$ decays at scales 
${\cal O}(m_b)$ is an effective five-quark field theory, which deals with 
local operators. Here $W^{\pm}$ and $t$ do not appear explicitly as dynamical 
fields and the local operators are built out of lighter flavours only. The 
effects of $W^{\pm}$ and $t$ are present only in the short-distance Wilson 
coefficients of these operators.

The purpose of this section is to clarify the relation between these
two approaches to describe $B$ decays and in particular to state explicitly 
what is meant by the $\bar b\to\bar s$ penguin amplitude in 
(\ref{ampl-char}), and similarly by the $\bar b\to\bar d$ penguin 
amplitude contributing to the decay $B^+\to K^+\overline{K^0}$, which will
play an important role in Section~\ref{pen-hunt}. To this end, let us recall 
the effective Hamiltonian for $\Delta B=+1$ decays by concentrating on the 
$\bar b \to \bar s$ penguin transitions. The $\bar b \to \bar d$ case 
can be analysed in the same manner with the obvious replacement $s \to d$ 
in the formulae given below. We have \cite{BJLW}
\begin{equation}\label{heff}
{\cal H}_{{\rm eff}} = \frac{G_{\rm F}}{\sqrt{2}}\left[ 
\lambda_u^{(s)}\left(C_1(\mu) Q_1^u + C_2(\mu) Q_2^u\right) 
+\lambda_c^{(s)}
\left(C_1(\mu) Q_1^c + C_2(\mu) Q_2^c\right)
-\lambda_t^{(s)} \sum^{6}_{i=3} 
C_i(\mu) Q_i \right],
\end{equation}
where $\mu$ is the renormalization scale ${\cal O}(m_b)$, 
$\lambda_i^{(s)}\equiv V_{is} V^{*}_{ib}$, and
\begin{equation}\label{O1} 
Q_1^c = (\bar c_{\alpha} s_{\beta})_{{\rm V-A}}\;(\bar b_{\beta} 
c_{\alpha})_{{\rm V-A}}\,,
~~~~~~Q_2^c = (\bar c s)_{{\rm V-A}}\;(\bar b c)_{{\rm V-A}} 
\end{equation}
\begin{equation}\label{O1u} 
Q_1^u = (\bar u_{\alpha} s_{\beta})_{{\rm V-A}}\;(\bar b_{\beta} 
u_{\alpha})_{{\rm V-A}}\,,
~~~~~~Q_2^u = (\bar u s)_{{\rm V-A}}\;(\bar b u)_{{\rm V-A}} 
\end{equation}
\begin{equation}\label{O2}
Q_3 = (\bar b s)_{{\rm V-A}}\sum_{q=u,d,s,c,b}(\bar qq)_{{\rm V-A}}\,,~~~~~~   
Q_4 = (\bar b_{\alpha} s_{\beta})_{{\rm V-A}}\sum_{q=u,d,s,c,b}(\bar 
q_{\beta} q_{\alpha})_{{\rm V-A}} 
\end{equation}
\begin{equation}\label{O3}
Q_5 = (\bar b s)_{{\rm V-A}} \sum_{q=u,d,s,c,b}(\bar qq)_{{\rm V+A}}\,,~~~~~  
Q_6 = (\bar b_{\alpha} s_{\beta})_{{\rm V-A}}\sum_{q=u,d,s,c,b}
(\bar q_{\beta} q_{\alpha})_{{\rm V+A}},
\end{equation}
with $\alpha$ and $\beta$ denoting colour indices. Here $Q^{u,c}_{1,2}$
are the current--current operators and the $Q_i$ (with $i=3,\ldots,6$) the QCD 
penguin operators. We do not show the electroweak penguin operators, which
can be found for instance in \cite{bf-rev}. A discussion similar to the one
presented below can also be made for the latter operators \cite{defan}. 
Since quark charges enter in the electroweak penguin operators, they exhibit, 
however, a different isospin structure as the QCD penguin operators.

The explicit derivation of (\ref{heff}) can be found in an appendix 
of \cite{BJLW}. It involves in particular the matching of the full
six-quark theory containing $W^{\pm}$ with the effective theory, in
which $W^{\pm}$ and the top quark do not appear as dynamical fields: the 
operators $Q_i$ do not involve the top quark and are built out of $u$, $c$, 
$d$, $s$ and $b$ quarks only.

\begin{figure}
\epsfysize=4.5cm
\centerline{\epsffile{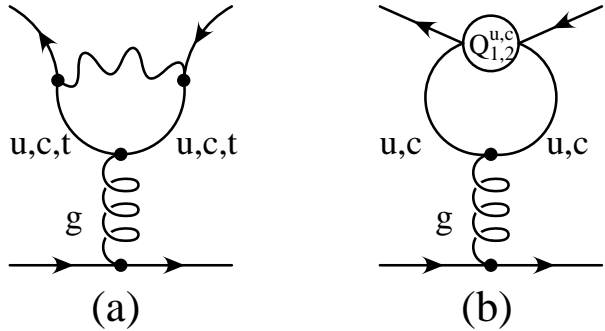}}
\caption{Penguin diagrams in the full (a) and effective (b) theory.
$Q^{u,c}_{1,2}$ denotes operator insertions.}\label{fig:ping}
\end{figure}

Let us now see how the penguin diagrams with internal $W^\pm$, $t$, $c$
and $u$ exchanges, in Fig.\ 1 (a), are represented in the effective 
Hamiltonian (\ref{heff}). The effect of the penguin diagrams with internal 
$W^\pm$ and top quark exchanges is obviously represented by the last term: 
the coefficients $C_{3-6}$ depend on $x_t\equiv m^2_t/M_W^2$, and this 
dependence is represented by the well-known Inami--Lim function $E(x_t)$ 
\cite{IL}. On the other hand, there is no trace in (\ref{heff}) of the 
penguin diagrams with internal $W^\pm$, $u$ and $c$ exchanges. In particular, 
the coefficients $C_{1,2}$ do not carry any information on these penguin 
topologies. Indeed, as explicitly demonstrated in \cite{BJLW}, the penguin 
diagrams with internal $u$ and $c$ quarks and with the $W$ propagator have 
been cancelled in the process of matching by similar diagrams, where the $W$ 
propagators have been replaced by local operators, as shown in Fig.\ 1 (b).
Strictly speaking, there remains a renormalization scheme dependent constant, 
which is added with the help of the unitarity of the CKM matrix to the 
function $E(x_t)$ in $C_{3-6}$. This constant is equal to $-2/3$ in the NDR 
scheme, and vanishes in the HV scheme \cite{BJLW}. A discussion of these 
features and applications to non-leptonic $B$ decays can also be found 
in \cite{rf}.

The fact that there is no trace of penguin diagrams with internal
$u$ and $c$ quarks in (\ref{heff}) is consistent with the general structure
of the operator product expansion combined with renormalization
group methods. At scales $\mu_b={\cal O}(m_b)$, the effect of quarks
with $m_i<\mu_b$ is absent in the Wilson coefficients and can only
be found in the matrix elements of the operators $Q_i$.

The $\bar b\to\bar s$ QCD penguin amplitude $P^{(s)}$ contributing to 
$B^+\to\pi^+K^0$ can be decomposed as follows:
\begin{equation}\label{ps}
P^{(s)} = \lambda_u^{(s)} P_u^{(s)}+\lambda_c^{(s)}  
P_c^{(s)}+\lambda_t^{(s)} P_t^{(s)}.
\end{equation}
Analogously, the $\bar b\to\bar d$ QCD penguin amplitude $P^{(d)}$ 
contributing to $B^+\to K^+\overline{K^0}$ can be written as
\begin{equation}\label{pd-def}
P^{(d)}=\lambda_u^{(d)} P_u^{(d)}+ 
\lambda_c^{(d)}  P_c^{(d)}+\lambda_t^{(d)} P_t^{(d)}. 
\end{equation}
Note that similar expressions hold also for the electroweak penguin amplitudes.
The strong amplitudes $P_q^{(s)}$ and $P_q^{(d)}$ ($q\in\{u,c,t\}$) in 
(\ref{ps}) and (\ref{pd-def}) are related to each other by interchanging 
all $d$ and $s$ quarks, i.e.\ through the so-called $U$ spin of the $SU(3)$ 
flavour symmetry of strong interactions. Let us focus in the following 
discussion on the decay $B^+\to\pi^+K^0$. In the formal operator language 
presented above, the meaning of  $P_u^{(s)}$, $ P_c^{(s)}$ and $ P_t^{(s)}$ 
is simply as follows:
\begin{equation}\label{psu}
P_u^{(s)}=\frac{G_{\rm F}}{\sqrt{2}}\left[C_1(\mu)\,\langle K^0\pi^+|
Q_1^u(\mu)|B^+\rangle_{\rm P}+\,C_2(\mu)\,\langle K^0\pi^+|Q_2^u(\mu)
|B^+\rangle_{\rm P}\right]
\end{equation}
\begin{equation}\label{psc}
P_c^{(s)}=\frac{G_{\rm F}}{\sqrt{2}}\left[C_1(\mu)\,\langle K^0\pi^+|
Q_1^c(\mu)|B^+\rangle_{\rm P} + \,
C_2(\mu)\,\langle K^0\pi^+|Q_2^c(\mu)|B^+\rangle_{\rm P}\right]
\end{equation}
\begin{equation}\label{pst}
P_t^{(s)}=-\,\frac{G_{\rm F}}{\sqrt{2}}\sum^{6}_{i=3} 
C_i(\mu)\,\langle K^0\pi^+|Q_i(\mu)|B^+\rangle .
\end{equation}
Here $\langle K^0\pi^+|Q_{1,2}^u|B^+\rangle_{\rm P}$ and $\langle K^0\pi^+|
Q_{1,2}^c|B^+\rangle_{\rm P}$ denote hadronic matrix elements with insertions 
of the current--current operators $Q_{1,2}^u$ and $Q_{1,2}^c$ into penguin 
diagrams with internal $u$ and $c$ quark exchanges. The importance of such 
diagrams in connection with certain strategies for CKM determinations has 
been pointed out for the first time in \cite{PAP0}, although such 
contributions have been considered already a long time ago in a different 
context \cite{gh}. Recently the importance of 
$\langle Q_{1,2}^c\rangle_{\rm P}$, in particular in connection 
with the CLEO data and the determination of the angle $\alpha$ through the CP 
asymmetry in $B_d\to \pi^+\pi^-$, has also been emphasized by the authors 
of \cite{ital}, who have named them ``charming penguins''.

Let us next observe that the matrix elements of the penguin operators
$Q_{3-6}$ do not carry the subscript P. Indeed, what is meant by
$P_t^{(s)}$ in the literature are hadronic matrix elements with insertions 
of the penguin operators not only into the penguin diagrams, but also into 
other topologies, in particular tree diagrams. 

At this point, it should be stressed that the current--current operators 
$Q^{c}_{1,2}$ and $Q^{u}_{1,2}$ contribute to the mode $B^+\to\pi^+K^0$
only through the penguin topologies discussed above, and through annihilation 
topologies. The latter, which will be discussed in more detail in the
following section, are described by the $A^{(s)}$ amplitude in 
(\ref{ampl-char}) and are only due to the $Q^{u}_{1,2}$ operators 
\cite{neubert}. Such annihilation processes are absent in the case of 
$B^0_d\to\pi^-K^+$. However, in contrast to $B^+\to\pi^+K^0$, this decay 
receives also contributions from hadronic matrix elements of the 
$Q^{u}_{1,2}$ operators with insertions into tree-diagram-like topologies, 
which are represented in (\ref{ampl-neut}) by the amplitude ${\cal T}^{(s)}$.

We hope that this discussion shows the relation between the formal 
operator method and the more phenomenological picture used by CP asymmetry 
practitioners. Simultaneously, this discussion demonstrates that attributing 
the usual Feynman diagrams with full $W$ propagators and internal $t$, $c$ 
and $u$ quarks to $P_t^{(s)}$, $P_c^{(s)}$ and $P_u^{(s)}$, respectively -- 
although roughly correct -- does not fully describe what happens at scales
${\cal O}(m_b)$ and may sometimes be confusing.

Let us next have a closer look at the general structure of the 
$\bar b\to\bar q$ $(q\in\{s,d\})$ QCD penguin amplitudes $P^{(s)}$ and 
$P^{(d)}$. Employing the unitarity of the CKM matrix and making furthermore 
use of the Wolfenstein parametrization \cite {wolf}, we arrive at
\begin{equation}\label{bspenamp}
P^{(s)}=-\left(1-\frac{1}{2}\lambda^2\right)\,\lambda^2\,A\,\left[1-
\Delta P^{(s)}\,+\,\left(\frac{\lambda^2\,R_b}{1-\lambda^2/2}\right)
\,e^{i\gamma}\right]\left|P_{tu}^{(s)}\right|\,e^{i\delta_{tu}^{(s)}}
\end{equation}
\begin{equation}\label{bdpenamp}
P^{(d)}=A\,\lambda^3\,R_t\,\left[e^{-i\beta}-\frac{1}{R_t}\Delta P^{(d)}
+{\cal O}(\lambda^4)\right]
\left|P_{tu}^{(d)}\right|\,e^{i\delta_{tu}^{(d)}},
\end{equation}
where the notation
\begin{equation}\label{q1q2}
P_{q_1q_2}^{(q)}\equiv\left|P_{q_1q_2}^{(q)}\right|\,
e^{i\delta_{q_1q_2}^{(q)}}\equiv P_{q_1}^{(q)}-P_{q_2}^{(q)}
\end{equation}
has been introduced, and 
\begin{equation}\label{DelP}
\Delta P^{(q)}\equiv\left|\Delta P^{(q)}\right|\,e^{i\delta_{\Delta P}^{(q)}}
\equiv\frac{P^{(q)}_{cu}}{P^{(q)}_{tu}}=\frac{P_c^{(q)}-P_u^{(q)}}{P_t^{(q)}-
P_u^{(q)}}
\end{equation}
describes the contributions of penguins with up and charm quarks running as 
virtual particles in the loops. The present status of the relevant CKM 
factors in (\ref{bspenamp}) and (\ref{bdpenamp}) is given by
\begin{equation}
A\equiv\frac{1}{\lambda^2}\left|V_{cb}\right|=0.81\pm0.06\,,\quad
R_b\equiv\frac{1}{\lambda}
\left|\frac{V_{ub}}{V_{cb}}\right|=0.36\pm0.08\,,\quad
R_t\equiv\frac{1}{\lambda}\left|\frac{V_{td}}{V_{cb}}\right|={\cal O}(1)\,,
\end{equation}
where $\lambda\equiv|V_{us}|=0.22$. Strategies to fix these parameters have 
recently been reviewed in \cite{bf-rev}. 

At this point, it should be emphasized that whereas the $P^{(q)}$ amplitudes
are $\mu$ and renormalization scheme independent, this is not the
case for the different contributions in (\ref{ps}) and (\ref{pd-def}). 
This is evident, if one inspects equations (\ref{psu})--(\ref{pst}). To 
this end, one can simply evaluate the penguin-like matrix elements of 
$Q^{u,c}_{1,2}$ in a perturbative framework to find that their $\mu$ 
dependences, related to the mixing between $Q^{u,c}_{1,2}$ and the penguin 
operators, cannot be cancelled by the $\mu$ dependence of $C_{1,2}(\mu)$. 
Similarly, the non-logarithmic terms in these matrix elements are 
renormalization scheme dependent, and this scheme dependence cannot be 
cancelled by the scheme dependence of $C_{1,2}(\mu)$. On the other hand, 
$P^{(q)}_{tu}$ and $\Delta P^{(q)}$ are $\mu$ independent and renormalization 
scheme independent, because these dependences cancel in (\ref{q1q2}). 
Consequently, $\Delta P^{(q)}$ is a physical quantity and can therefore be 
determined experimentally, as we will see in Section~\ref{pen-hunt}.

Formulae (\ref{bspenamp}) and (\ref{bdpenamp}) can already be found 
in the literature \cite{PAP0,rev}. In the case of the $P^{(s)}$ amplitude
(\ref{bspenamp}), we have kept terms of ${\cal O}(\lambda^4)$, which have 
been neglected in these papers, and also in the bound on the CKM angle 
$\gamma$ derived in \cite{fm2}. The highly CKM-suppressed $\lambda^2\,R_b\,
e^{i\gamma}={\cal O}(0.02)$ phase factor may lead to direct CP violation 
in the decay $B^+\to\pi^+ K^0$. Model calculations performed at the 
perturbative quark level indicate that $\Delta P^{(s)}$ is not close to 1, 
i.e.\ that the large CKM suppression of the $e^{i\gamma}$ term in 
(\ref{bspenamp}) is not compensated, and give CP asymmetries of at most a 
few percent \cite{rf,gh,calcs}. However, as was pointed out recently 
\cite{gewe}--\cite{atso}, rescattering effects of the kind $B^+\to\{\pi^0K^+\}
\to\pi^+K^0$ may lead to CP asymmetries as large as ${\cal O}(10\%)$, 
and represent an important limitation of the theoretical accuracy of 
the ``original'' bounds on $\gamma$ arising from the $B^\pm\to\pi^\pm K$ and
$B_d\to\pi^\mp K^\pm$ decays that were proposed in \cite{fm2}. Let us 
have a closer look at these final-state interaction effects, which are
closely related to penguin topologies, in the following section.

\section{Isospin Relations and the Connection Between\\ 
Rescattering Processes and Penguin Topologies}\label{pen-zoo}
As we have already noted, the decays $B^+\to\pi^+K^0$, $B^0_d\to\pi^-K^+$
and their charge conjugates provide a fertile ground to probe the CKM angle
$\gamma$ \cite{PAPIII}--\cite{groro}. In this context, one makes use of 
the $SU(2)$ isospin symmetry of strong interactions, which appears -- in 
contrast to the $SU(3)$ flavour symmetry -- to be a very good and safe 
working assumption. It allows us to relate the QCD penguin amplitudes 
contributing to $B^+\to\pi^+K^0$ and $B^0_d\to\pi^-K^+$ to each other, 
yielding the following relations:
\begin{eqnarray}
A(B^+ \to \pi^+K^0)&\equiv&P\label{ampl-chariso}\\
A(B^0_d \to\pi^- K^+)&=&-\,\left[\,T+P+P_{\rm ew}\,
\right]\label{ampl-neutiso}\,,
\end{eqnarray}
where the $B^+ \to \pi^+K^0$ decay amplitude {\it defines} the 
$\bar b\to\bar s$ ``penguin'' amplitude $P$, the quantity 
$P_{\rm ew}\equiv c_u \,{\cal P}_{{\rm EW}}^{(s){\rm C}}\,-\,c_d\, 
P_{{\rm EW}}^{(s){\rm C}}$ is due to ``colour-suppressed'' electroweak 
penguins, and the generalized ``colour-allowed'' $\bar b\to\bar uu\bar s$ 
``tree'' amplitude takes the form
\begin{equation}\label{tree}
T = e^{i\gamma} e^{i\delta_T} |T|\,,
\end{equation}
where $\delta_T$ is a CP-conserving strong phase. In order to express 
(\ref{ampl-char}) and (\ref{ampl-neut}) as in (\ref{ampl-chariso}) and 
(\ref{ampl-neutiso}) with the help of the $SU(2)$ isospin symmetry, special 
care has to be taken. In particular, the ``tree'' amplitude $T$ has to be 
defined in a proper way. The point is that an application of the isospin 
symmetry requires that we replace all $u$ and $d$ quarks in the 
$B^+\to\pi^+K^0$ decay processes by $d$ and $u$ quarks, respectively, in
order to relate them to those of the transition $B^0_d\to\pi^-K^+$. While 
such a replacement is straightforward in the case of 
\begin{equation}\label{psc-iso}
{\cal P}_c^{(s)}=\frac{G_{\rm F}}{\sqrt{2}}\left[C_1(\mu)\,\langle K^+\pi^-|
Q_1^c(\mu)|B^0_d\rangle_{\rm P} + \,
C_2(\mu)\,\langle K^+\pi^-|Q_2^c(\mu)|B^0_d\rangle_{\rm P}\right]=P_c^{(s)}
\end{equation}
\begin{equation}\label{pst-iso}
{\cal P}_t^{(s)}=-\,\frac{G_{\rm F}}{\sqrt{2}}\sum^{6}_{i=3} 
C_i(\mu)\,\langle K^+\pi^-|Q_i(\mu)|B^0_d\rangle=P_t^{(s)},
\end{equation}
at first sight a problem shows up in the case of
\begin{equation}\label{psu-tilde}
{\cal P}_u^{(s)}=\frac{G_{\rm F}}{\sqrt{2}}\left[C_1(\mu)\,\langle K^+\pi^-|
Q_1^u(\mu)|B^0_d\rangle_{\rm P}+\,C_2(\mu)\,\langle K^+\pi^-|Q_2^u(\mu)
|B^0_d\rangle_{\rm P}\right],
\end{equation}
since isospin symmetry implies the relation
\begin{eqnarray}
\lefteqn{P_u^{(s)}=\frac{G_{\rm F}}{\sqrt{2}}\left[C_1(\mu)\,\langle K^0\pi^+|
Q_1^u(\mu)|B^+\rangle_{\rm P}+\,C_2(\mu)\,\langle K^0\pi^+|Q_2^u(\mu)
|B^+\rangle_{\rm P}\right]}\nonumber\\
&&=-\,\frac{G_{\rm F}}{\sqrt{2}}\left[C_1(\mu)\,\langle K^+\pi^-|
Q_1^d(\mu)|B^0_d\rangle_{\rm P}+\,C_2(\mu)\,\langle K^+\pi^-|Q_2^d(\mu)
|B^0_d\rangle_{\rm P}\right].\label{psu-iso}
\end{eqnarray}
Here $Q_{1,2}^d$ can be obtained easily from (\ref{O1u}) by substituting $u$
through $d$ quarks. Consequently, isospin does not imply that $P_u^{(s)}$ is
equal to ${\cal P}_u^{(s)}$. The difference between these amplitudes can,
however, be absorbed in the proper definition of the $T$ amplitude in 
(\ref{ampl-neutiso}). It has to be defined as
\begin{eqnarray}
\lefteqn{T\equiv-\,\frac{G_{\rm F}}{\sqrt{2}}\,\lambda^{(s)}_u\biggl[
C_1(\mu) \langle K^+\pi^-|Q_1^u(\mu)|B^0_d\rangle_{\rm T} + 
C_2(\mu) \langle K^+\pi^-|Q_2^u(\mu)|B^0_d\rangle_{\rm T}}\nonumber\\
&&~~~~~~~+\Bigl\{C_1(\mu) \langle K^+\pi^-|Q_1^u(\mu)|B^0_d\rangle_{\rm P} + 
C_2(\mu) \langle K^+\pi^-|Q_2^u(\mu)|B^0_d\rangle_{\rm P}\nonumber\\
&&~~~~~~~-C_1(\mu) \langle K^+\pi^-|Q_1^d(\mu)|B^0_d\rangle -
C_2(\mu) \langle K^+\pi^-|Q_2^d(\mu)|B^0_d\rangle\Bigr\}\biggr]\label{Ts-def}
\end{eqnarray}
to arrive at (\ref{ampl-chariso}) and (\ref{ampl-neutiso}).

\begin{figure}
\begin{center}
\leavevmode
\epsfysize=4.5truecm 
\epsffile{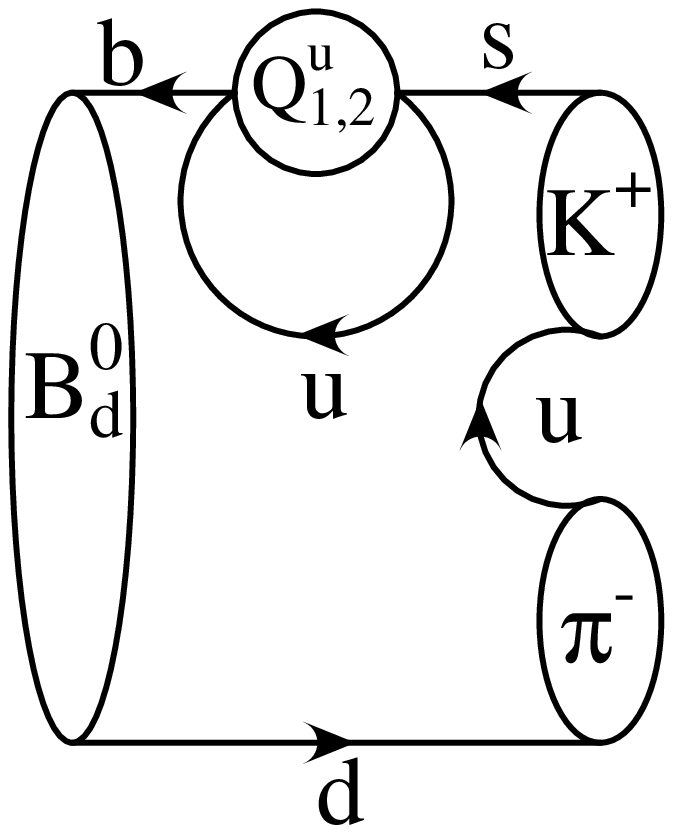} 
  \hspace*{1cm} \boldmath 
  \raisebox{2.3cm}{\large $-$}  \unboldmath \hspace*{1cm}
\epsfysize=4.5truecm 
\epsffile{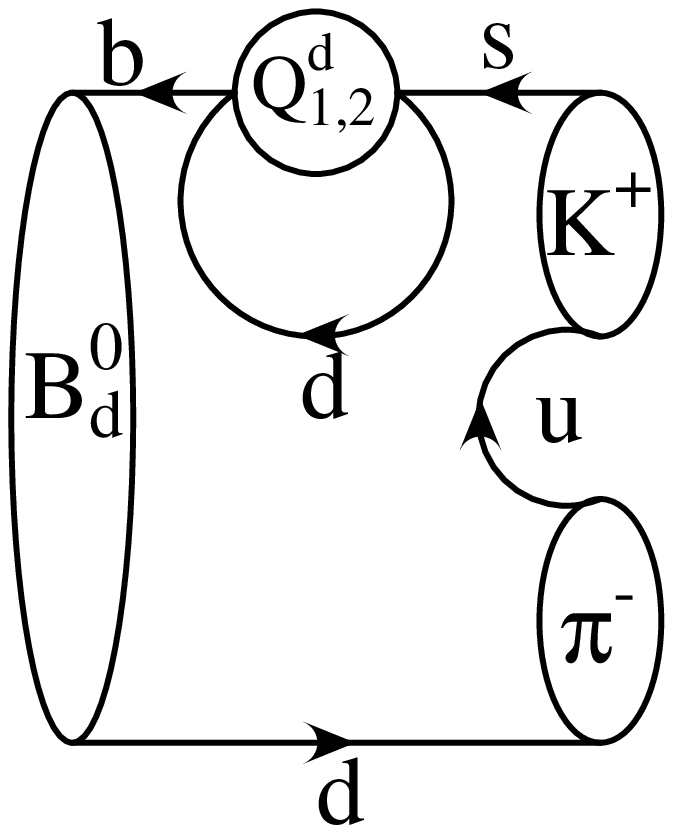}
\end{center}
\caption{Penguin topologies contributing to the 
amplitude $T$.}\label{fig:pen-diff}
\end{figure}

Equations (\ref{psc-iso}), (\ref{pst-iso}) and (\ref{psu-iso}) are completely
general and rely only on the isospin decomposition of the $B^+\to\pi^+K^0$
and $B^0_d\to\pi^-K^+$ decay amplitudes, which is performed explicitly by
using the Wigner--Eckart theorem in the appendix. The second term 
in (\ref{Ts-def}), which is required by this isospin decomposition, shows 
that $T$ is not only given by hadronic matrix elements with insertions 
of the $Q_{1,2}^u$ current--current operators into tree-diagram-like 
topologies, as na\"\i vely expected. The $Q_{1,2}^d$ operators contribute 
both through insertions into penguin topologies and through annihilation 
processes. The latter correspond to the annihilation amplitude $A^{(s)}$ 
appearing in (\ref{ampl-char}). Consequently, the term in curly brackets in
(\ref{Ts-def}) consists both of the difference of contributions from penguin 
topologies with internal up and down quarks, which is shown in 
Fig.~\ref{fig:pen-diff}, and of contributions from annihilation topologies. 

In order to address the question of rescattering effects in $B\to\pi K$
decays, let us first return to the expressions given in (\ref{bspenamp}) and 
(\ref{bdpenamp}). These formulae are completely general. In several previous 
analyses of penguin-induced $B$ decays, it was, however, assumed that penguin
processes are dominated by internal top quark exchanges, corresponding to 
$\Delta P^{(q)}=0$. An important implication of this special case would be 
that the $P^{(d)}$ penguin amplitude is proportional to the phase factor 
$e^{-i\beta}$, i.e.\ exhibits a very simple phase structure. As was pointed 
out in \cite{PAP0}, this feature is spoiled by penguin contributions with 
internal up and charm quarks, leading to sizeable values of $\Delta P^{(d)}$. 
Since the $e^{i\gamma}$ factor in (\ref{bspenamp}) is highly CKM-suppressed 
by $\lambda^2R_b={\cal O}(0.02)$, there is to a good approximation no 
non-trivial CP-violating phase present in the $\bar b\to\bar s$ penguin 
amplitude, provided the relation $|1-\Delta P^{(s)}|\gg\lambda^2R_b$ is 
fulfilled. Model calculations performed at the perturbative quark-level to 
estimate the penguin amplitudes $P_q^{(s)}$ ($q\in\{u,c,t\}$) indicate that 
this requirement is indeed satisfied, i.e.\ that the very large CKM 
suppression of the CP-violating phase factor in (\ref{bspenamp}) is not 
compensated, and that there is direct CP violation in $B^\pm\to\pi^\pm K$ 
of at most a few percent \cite{gh,calcs}. 

\begin{figure}
\begin{center}
\leavevmode
\epsfysize=5cm 
\epsffile{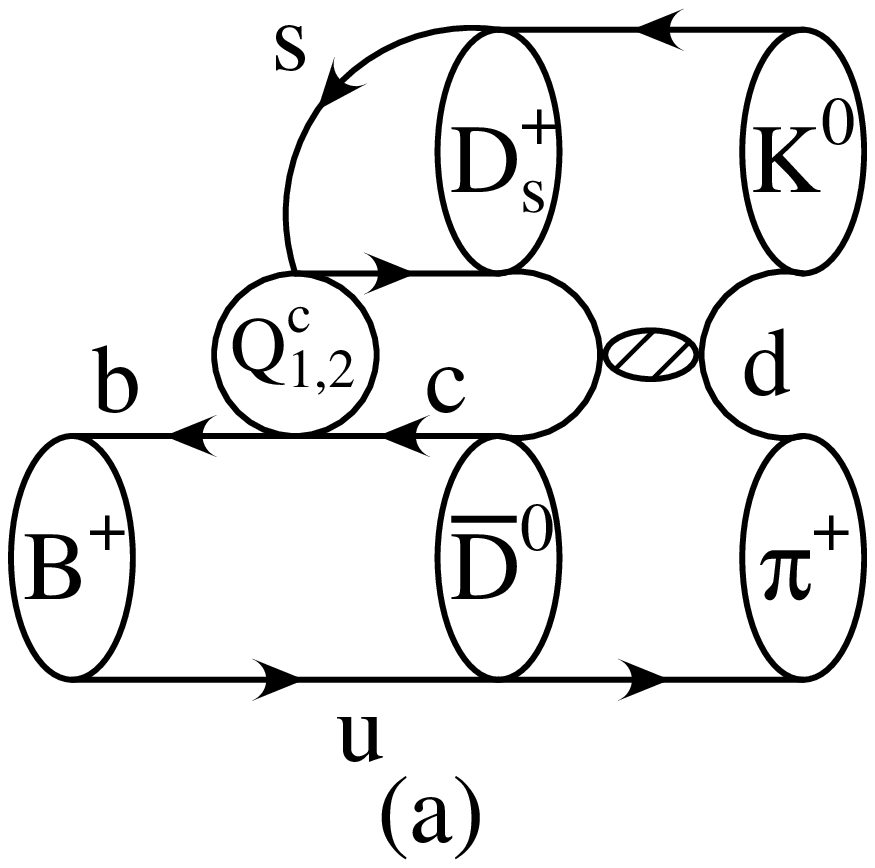} 
  \hspace*{1cm} \boldmath 
  \raisebox{2.3cm}{\large $\in$}  \unboldmath \hspace*{1cm}
\epsfysize=5cm 
\epsffile{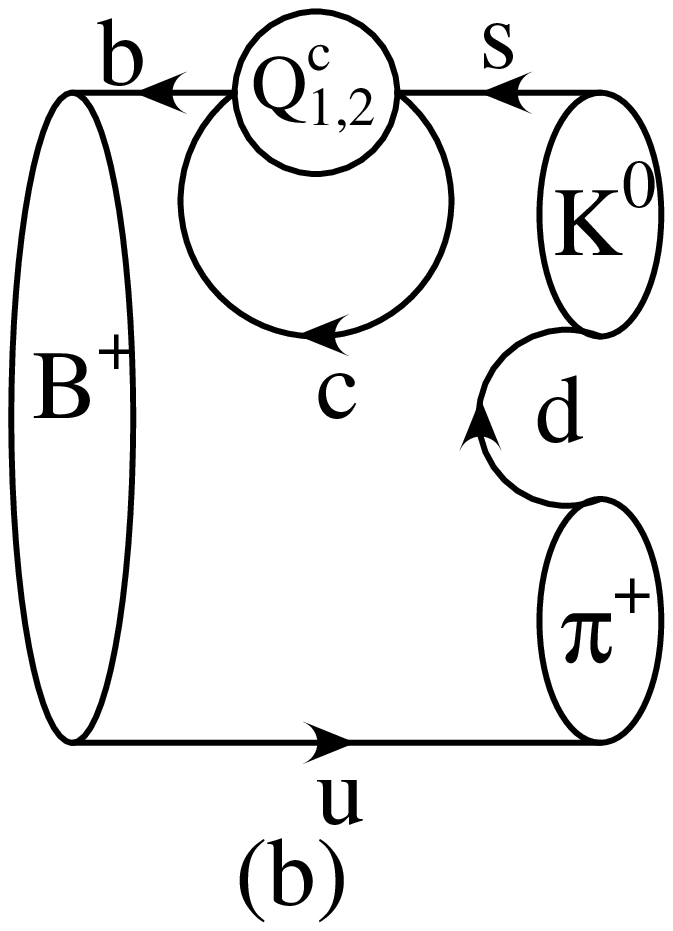}
\end{center}
\caption{Illustration of a rescattering processes of the kind 
$B^+\to\{\overline{D^0}D_s^+\}\to\pi^+K^0$ (a), which is contained in 
penguin topologies with 
internal charm quarks (b). $Q^c_{1,2}$ denotes an 
operator insertion.}\label{fig:pen-c}
\end{figure}

The QCD penguin amplitudes $P_q^{(d,s)}$ receive, however, also long-distance
contributions from rescattering processes, which can be divided into two
categories and are not included in these simple model calculations. Let us 
discuss these effects by focusing on the channel $B^+\to\pi^+K^0$. In the
first class of rescattering processes, we have to deal with decays such
as $B^+\to \overline{D^0}D_s^+$, which are caused by the $Q_{1,2}^{c}$ 
current--current operators through insertions into tree-diagram-like 
topologies, and may rescatter into the final state $\pi^+K^0$, 
i.e.\ we have 
\begin{equation}\label{c-rescatter}
B^+\to\{F_c^{(s)}\}\to\pi^+K^0,
\end{equation}
where $F_c^{(s)}\in\{\overline{D^0}D_s^+,\,\overline{D^0}D_s^{\ast+},\,
\overline{D^{\ast 0}}D_s^{\ast+},\,\ldots\}$. Here the dots include also
intermediate multibody states. These final-state interaction
effects are related to penguin topologies with internal charm quark 
exchanges, as can be seen in Fig.~\ref{fig:pen-c}, and are included in 
(\ref{bspenamp}) as long-distance contributions to the $P_c^{(s)}$ amplitude. 
They are expected to contribute significanlty to the magnitude of the 
$\bar b\to\bar s$ QCD penguin amplitude \cite{PAP0,ital}. 

\begin{figure}
\begin{center}
\leavevmode
\epsfysize=5cm 
\epsffile{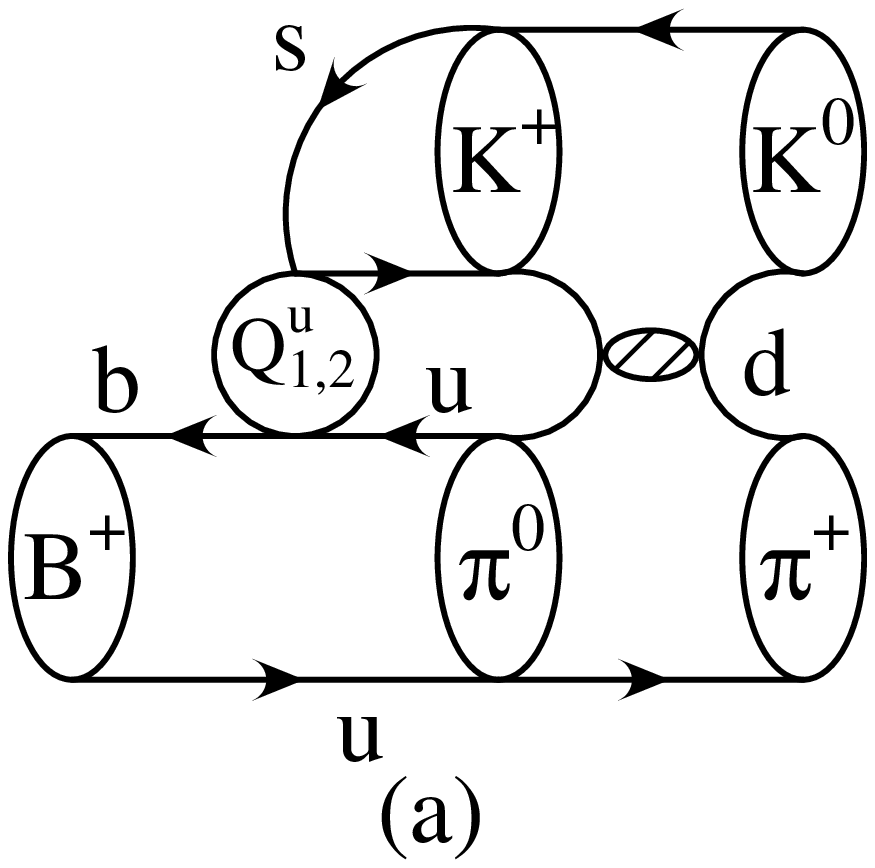} 
  \hspace*{1cm} \boldmath 
  \raisebox{2.3cm}{\large $\in$}  \unboldmath \hspace*{1cm}
\epsfysize=5cm 
\epsffile{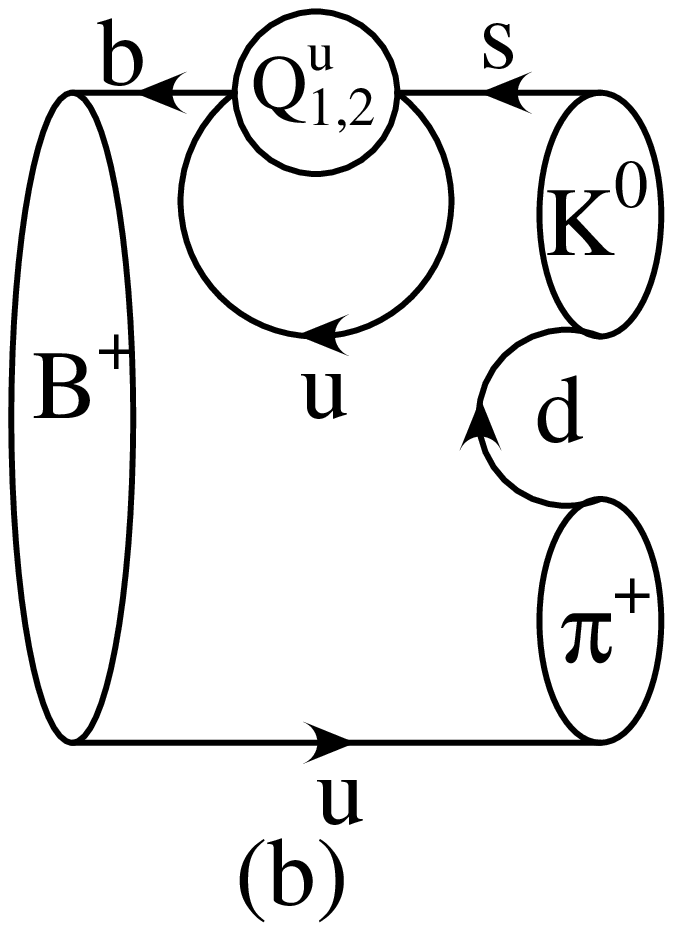}
\end{center}
\caption{Illustration of a rescattering processes of the kind 
$B^+\to\{\pi^0K^+\}\to\pi^+K^0$ (a), which is contained in 
penguin topologies with internal up
quarks (b). $Q^u_{1,2}$ denotes an 
operator insertion.}\label{fig:pen-u}
\end{figure}

In the second class of rescattering processes, channels of the kind $B^+\to
\pi^0K^+$, which receive $Q_{1,2}^u$ current--current operator contributions 
with insertions into tree-diagram-like topologies, rescatter into $\pi^+K^0$,
i.e.\
\begin{equation}\label{u-rescatter}
B^+\to\{F_u^{(s)}\}\to\pi^+K^0,
\end{equation}
where $F_u^{(s)}\in\{\pi^0K^+,\,\pi^0K^{\ast +},\,\rho^0K^{\ast +},\,
\ldots\}$. As can be seen in Fig.~\ref{fig:pen-u}, these final-state
interaction effects are related to penguin topologies with internal up quarks 
and appear in (\ref{bspenamp}) as long-distance contributions to the 
$P_u^{(s)}$ amplitude. Moreover, we also get contributions from the
rescattering processes (\ref{u-rescatter}) to the amplitude $A^{(s)}$ 
through annihilation topologies. Usually it is assumed that annihilation 
processes are suppressed relative to tree-diagram-like processes by a factor 
of $f_B/m_B$. However, this feature may no longer hold in the presence of 
rescattering effects \cite{neubert,Bloketal}, so that annihilation 
topologies may play a more important role than na\"\i vely expected. Model 
calculations \cite{Bloketal} based on Regge phenomenology typically give an 
enhancement of the ratio $|A^{(s)}|/|{\cal T}^{(s)}|$ from 
$f_B/m_B\approx0.04$ to ${\cal O}(0.2)$. Rescattering processes of this 
kind can be probed, e.g.\ by the $\Delta S$\,=\,0 decay $B_d\to K^+K^-$. 
A future stringent bound on BR$(B_d\to K^+K^-)$ at the level of $10^{-7}$ 
or lower may provide a useful limit on these rescattering effects 
\cite{groro}. The present upper bound obtained by the CLEO collaboration is 
$4.3\times10^{-6}$ \cite{cleo}. In order to take into account annihilation
processes in the $B^+\to\pi^+ K^0$ decay amplitude, we simply have to
perform the replacement $P_u^{(s)}\to P_u^{(s)}+\widetilde{A}^{(s)}$ in
(\ref{DelP}), yielding
\begin{equation}\label{DelPtilde}
\Delta\widetilde{P}^{(s)}\equiv\frac{P_c^{(s)}-P_u^{(s)}-
\widetilde{A}^{(s)}}{P_t^{(s)}-P_u^{(s)}-\widetilde{A}^{(s)}}\,,
\end{equation}
where
$\widetilde{A}^{(s)}$ is defined by $A^{(s)}\equiv V_{us}V_{ub}^\ast\, 
\widetilde{A}^{(s)}$.

The quantity $\Delta\widetilde{P}^{(s)}$ would not be close to 1 if 
rescattering processes of the type (\ref{c-rescatter}) played the dominant 
role in $B^+\to\pi^+K^0$, or if both (\ref{c-rescatter}) and 
(\ref{u-rescatter}) were similarly important. In the former case, 
$\Delta\widetilde{P}^{(s)}$ would carry a sizeable 
CP-conserving strong phase, whereas there would be a cancellation in 
(\ref{DelPtilde}) in the latter case, leading to $|\Delta\widetilde{P}^{(s)}|
\ll1$. On the other hand, $\Delta\widetilde{P}^{(s)}$ may be close to 1, if 
the final-state interactions arising from (\ref{u-rescatter}) would dominate 
$B^+\to\pi^+K^0$. In a recent attempt to calculate final-state 
interaction effects of the kind $B^+\to\{\pi^0K^+\}\to\pi^+K^0$ with the help 
of Regge phenomenology \cite{fknp}, it is found that such rescattering 
processes may in fact play a dominant role, i.e.\ 
$|{\cal P}_{uc}|/|{\cal P}_{tc}|={\cal O}(5)$, thereby leading to values of 
$|1-\Delta\widetilde{P}^{(s)}|$ as small as ${\cal O}(0.2)$. An important 
phenomenological implication of these rescattering effects would
be CP violation in $B^+\to\pi^+K^0$ as large as ${\cal O}(10\%)$. Similar
features are also found in a different approach to deal with final-state
interactions in $B\to\pi K$ decays \cite{gewe,neubert}.

Although the ``factorization'' hypothesis \cite{facto} is in general 
questionable, it may work reasonably well for the colour-allowed amplitude 
${\cal T}^{(s)}$ \cite{bjorken}. Since the intrinsic ``strength'' of 
decays such as $B^+\to\pi^0K^+$, representing the ``first step'' of the 
rescattering processes (\ref{u-rescatter}), is given by  
${\cal T}^{(s)}$, we may derive a ``plausible'' upper bound 
$\lambda^2R_b/|1-\Delta\widetilde{P}^{(s)}|
\mathrel{\hbox{\rlap{\hbox{\lower4pt\hbox{$\sim$}}}\hbox{$<$}}}0.15$, where
the recent CLEO data on the combined $B^\pm\to\pi^\pm K$ branching ratio 
\cite{cleo}, and the BSW form factors \cite{BSW} have been used to evaluate 
${\cal T}^{(s)}$ within ``factorization'' (see \cite{defan}). Note 
that the ratio $\lambda^2R_b/|1-\Delta\widetilde{P}^{(s)}|$ is typically 
one order of magnitude smaller, if rescattering processes do not play the 
dominant role in $B^+\to\pi^+ K^0$. 

In the following discussion we will not comment further on quantitative
estimates of rescattering effects. A reliable theoretical treatment is very 
difficult and requires knowledge of the dynamics of strong interactions 
that is unfortunately not available at present. We rather advocate to use
additional experimental information to obtain insights into final-state
interactions. Before turning to such strategies in Section~\ref{pen-hunt},
let us first point out an interesting feature of the rescattering processes. 
The ratio $R$ of the combined $B\to\pi K$ branching ratios introduced in 
(\ref{Def-R}) does not only imply constraints on the CKM angle $\gamma$, 
but also on the magnitude of the ``tree'' amplitude $T$. Since the present 
central value $R=0.65$ obtained by the CLEO collaboration \cite{cleo} is 
significanlty smaller than 1, these constraints are at the edge of 
compatibility with ``factorization'' \cite{fm2,defan}. In particular, larger 
values of $|T|$ are favoured by the CLEO data. However, as we have seen in 
(\ref{Ts-def}), the properly defined amplitude $T$ is not just a 
``colour-allowed tree amplitude'' as ${\cal T}^{(s)}$, but actually receives 
also contributions from penguin and annihilation topologies. Consequently, 
if the rescattering effects characterized by (\ref{u-rescatter}) and related 
to such topologies play in fact an important role, the value of $|T|$ could 
in principle be shifted significantly from its ``factorized'' value. In 
particular, the small central value of $R=0.65$ may indicate already that 
$|T|$ is enhanced considerably by final-state interactions, and it may well 
be possible that future measurements will stabilize around this na\"\i vely 
small value \cite{defan}. 
 
The bounds on $\gamma$ derived in \cite{fm2} are constructed in such a way 
that they do {\it not} depend on $|T|$. Generalized bounds, making not only
use of the ratio $R$ of the combined $B\to\pi K$ branching ratios, but 
also of the ``pseudo-asymmetry'' $A_0$, which is defined by
\begin{equation}
A_0\equiv\frac{\mbox{BR}(B^0_d\to\pi^-K^+)-\mbox{BR}(\overline{B^0_d}\to
\pi^+K^-)}{\mbox{BR}(B^+\to\pi^+K^0)+\mbox{BR}(B^-\to\pi^-\overline{K^0})},
\end{equation}
were derived in \cite{defan}. They are due to the fact that the amplitude
relations (\ref{ampl-chariso}) and (\ref{ampl-neutiso}) imply the following
minimal value of $R$:
\begin{equation}\label{Rmin}
R_{\rm min}=\kappa\,\sin^2\gamma\,+\,
\frac{1}{\kappa}\left(\frac{A_0}{2\,\sin\gamma}\right)^2,
\end{equation}
where rescattering and electroweak penguin effects are included through the 
parameter $\kappa$, which is given by
\begin{equation}\label{kappa-def}
\kappa=\frac{1}{w^2}\left[\,1+2\,(\epsilon\,w)\cos\Delta+
(\epsilon\,w)^2\,\right]
\end{equation}
with
\begin{equation}
w=\sqrt{1+2\,\rho\,\cos\theta\cos\gamma+\rho^2}.
\end{equation}
Note that no approximations were made in order to derive (\ref{Rmin}).
The quantities $\rho$ and $\epsilon$ measure the ``strength'' of the
rescattering and electroweak penguin effects, respectively, and can be 
expressed as
\begin{equation}\label{rho-def}
\rho\,e^{i\theta}=\frac{\lambda^2R_b}{1-\lambda^2/2}\left(\frac{1}{1-\Delta
\widetilde{P}^{(s)}}\right)=\frac{\lambda^2R_b}{1-\lambda^2/2}\left[1-\left(
\frac{P_{uc}^{(s)}+\widetilde{A}^{(s)}}{P_{tc}^{(s)}}\right)\right]
\end{equation}
and 
\begin{equation}
\epsilon\equiv\frac{|P_{\rm ew}|}{\sqrt{\langle|P|^2\rangle}}\,,
\end{equation}
where
\begin{equation}
\langle|P|^2\rangle\equiv\frac{1}{2}\left(|P|^2+|\overline{P}|^2\right).
\end{equation}
The phase $\Delta$ in (\ref{kappa-def}) is given by the difference of 
CP-conserving strong phases of the QCD and electroweak penguins. Since the 
values of the CKM angle $\gamma$ implying $R_{\rm min}>R_{\rm exp}$, where 
$R_{\rm exp}$ denotes the experimentally determined value of $R$, are 
excluded, we obtain an allowed range for $\gamma$. For values of $R$ as 
small as 0.65, which is the central value of present CLEO data \cite{cleo}, 
a large region around $\gamma=90^\circ$ is excluded. As soon as a 
non-vanishing experimental result for $A_0$ has been established, also an 
interval around $\gamma=0^\circ$ and $180^\circ$ can be ruled out, while 
the impact on the excluded region around $90^\circ$ is rather small 
\cite{defan}. The ``original'' bounds derived in \cite{fm2} correspond to
$\rho=\epsilon=0$ and were obtained without using information provided by
the ``pseudo-asymmetry'' $A_0$, i.e.\ $R_{\rm min}^{(0)}=\sin^2\gamma$. 
The impact of rescattering and electroweak penguin effects on the constraints
on $\gamma$ implied by (\ref{Rmin}) was analysed in great detail in 
\cite{defan,rf-FSI}. As was pointed out in these papers, additional 
experimental data on the decay $B^+\to K^+\overline{K^0}$ and its charge 
conjugate allows us to take into account rescattering effects in these 
bounds {\it completely}. Let us have a closer look at another strategy to
accomplish this task in the following section.

\boldmath
\section{Penguin Hunting with $B_{u,d}\to K\overline{K}$ and\\
$B^\pm\to\pi^\pm K$ Decays}\label{pen-hunt}
\unboldmath
The penguin-induced $\bar b\to\bar d$ mode $B_d^0\to K^0\overline{K^0}$, which 
is governed by a decay amplitude as the one given in (\ref{bdpenamp}), was
proposed frequently in the literature as a probe for new physics. This
mode would indeed provide a striking signal of physics beyond the Standard
Model, if QCD penguin processes were dominated by internal top quarks. In
that case, the weak decay and $B^0_d$--$\overline{B^0_d}$ mixing phases would
cancel, implying vanishing CP violation in $B_d\to K^0\overline{K^0}$
(for a detailed discussion, see \cite{rev}). As we have seen in the previous 
section, this feature is, however, spoiled by penguin topologies with internal
up and charm quark exchanges, i.e.\ by the $\Delta P^{(d)}$ term 
\cite{BdK0K0bar}. Interestingly, the CP-violating asymmetries
induced by these contributions allow a determination of the CKM angle 
$\alpha$ that does not suffer from penguin uncertainties, if one relates
$B_d\to K^0\overline{K^0}$ and $B_d\to\pi^+\pi^-$ to each other with the help 
of the $SU(3)$ flavour symmetry of strong interactions \cite{PAPII}. We shall 
briefly come back to the issue of new-physics effects in 
$B_d\to K^0\overline{K^0}$ in Section~\ref{new-phys}.

In the present section, we point out that $B_d\to K^0\overline{K^0}$ and its 
spectator-quark isospin partner $B^\pm\to K^\pm K$ play an important role to 
obtain insights into the dynamics of penguin and rescattering processes. 
As in the case of the $B\to\pi K$ decays discussed above, the 
observables that will be available first are probably the combined branching 
ratios
\begin{eqnarray}
\mbox{BR}(B_d\to K^0\overline{K^0})&\equiv&\frac{1}{2}\left[\mbox{BR}(B_d^0\to 
K^0\overline{K^0})+\mbox{BR}(\overline{B_d^0}\to K^0\overline{K^0})
\right]\\
\mbox{BR}(B^\pm\to K^\pm K)&\equiv&\frac{1}{2}\left[\mbox{BR}(B^+\to 
K^+\overline{K^0})+\mbox{BR}(B^-\to K^-K^0)\right].
\end{eqnarray}
At present, only the bounds BR$(B_d\to K^0\overline{K^0})<1.7\times10^{-5}$ 
and BR$(B^\pm\to K^\pm K)<2.1\times10^{-5}$ are available from the CLEO
collaboration, while the combined $B^\pm\to\pi^\pm K$ branching ratio
\begin{eqnarray}
\mbox{BR}(B^\pm\to\pi^\pm K)&\equiv&\frac{1}{2}
\left[\mbox{BR}(B^+\to\pi^+ K^0)+\mbox{BR}(B^-\to\pi^-\overline{K^0})\right]
\nonumber\\
&=&\left(2.3^{+1.1}_{-1.0}\pm0.3\pm0.2\right)\times 10^{-5}\label{BR-com}
\end{eqnarray}
has already been measured \cite{cleo}.

In order to obtain insights into the dynamics of penguin and final-state
interaction processes, the ratio 
\begin{eqnarray}
H&\equiv& R_{SU(3)}^2\left(\frac{1-\lambda^2}{\lambda^2}\right)
\frac{\mbox{BR}(B^\pm\to K^\pm K)}{\mbox{BR}(B^\pm\to\pi^\pm K)}\nonumber\\
&=&\frac{R_t^2-2\,R_t\,
|\Delta P|\,\cos\delta_{\Delta P}\,\cos\beta+|\Delta P|^2}{1+2\,\lambda^2\,
\widetilde{R}_b\,\cos\gamma+\lambda^4\widetilde{R}_b^2-2\,|\Delta P|\,
\cos\delta_{\Delta P}\left(1+\lambda^2\widetilde{R}_b\cos\gamma\right)+
|\Delta P|^2}\label{H-def}
\end{eqnarray}
plays a key role. Here we have used the $SU(3)$ flavour symmetry of strong 
interactions to relate $\widetilde{P}^{(d)}$ to $\widetilde{P}^{(s)}$: 
\begin{equation}\label{su3-2}
\Delta\widetilde{P}^{(s)}=\Delta\widetilde{P}^{(d)}\equiv|\Delta P|
\,e^{i\delta_{\Delta P}},
\end{equation}
and $\widetilde{R}_b\equiv R_b/(1-\lambda^2/2)$. The quantity
\begin{equation}
R_{SU(3)}=\frac{M_B^2-M_\pi^2}{M_B^2-M_K^2}\,
\frac{F_{B\pi}(M_K^2;0^+)}{F_{BK}(M_K^2;0^+)}
\end{equation}
describes factorizable $SU(3)$ breaking, where $F_{B\pi}(M_K^2;0^+)$ and 
$F_{BK}(M_K^2;0^+)$ are form factors parametrizing the hadronic quark-current
matrix elements 
$\langle\pi|(\bar b d)_{{\rm V-A}}|B\rangle$ and $\langle
K|(\bar b s)_{{\rm V-A}}|B\rangle$, respectively. Using, for example, the
model of Bauer, Stech and Wirbel \cite{BSW}, we have 
$R_{SU(3)}={\cal O}(0.7)$. At present, there is unfortunately no reliable 
approach available to deal with non-factorizable $SU(3)$ breaking. Since 
already the factorizable corrections are significant, we expect that 
non-factorizable $SU(3)$ breaking may also lead to sizeable effects. 

Let us assume for the moment that $H$ is found experimentally to be different
from~1. Introducing then the quantity
\begin{equation}\label{h-def}
h\equiv\left|\frac{H-R_t\,\cos\beta}{1-H}\right|
\end{equation}
and keeping the strong phase $\delta_{\Delta P}$ in (\ref{H-def}) as an
unknown, free parameter, we arrive at the following range for $|\Delta P|$:
\begin{equation}\label{DP-bound}
\left|h-\sqrt{h^2+\left(\frac{H-R_t^2}{1-H}\right)}\right|\leq|\Delta P|\leq
h+\sqrt{h^2+\left(\frac{H-R_t^2}{1-H}\right)}\,,
\end{equation}
where terms of ${\cal O}(\lambda^2\,\widetilde{R}_b\,\cos\gamma)$ have
been neglected. This range shrinks to
\begin{equation}
|\Delta P|=\sqrt{\frac{H-R_t^2}{1-H}}
\end{equation}
for $h=0$, corresponding to $\cos\beta=H/R_t$. In the special case $H=1$, 
which has been excluded in the formulae given above, we have on the other hand 
\begin{equation}
|\Delta P|\geq\frac{1}{2}\left|\frac{1-R_t^2}{1-R_t\,\cos\beta}\right|\,.
\end{equation}

In order to constrain $|\Delta P|$ following this approach, both the CKM angle
$\beta$, which is measured in a clean way through mixing-induced CP violation 
in $B_d\to J/\psi\, K_{\rm S}$, and the parameter $R_t$, fixing one side 
of the unitarity triangle, have to be known. The most promising ways to 
determine $R_t$ are the ratio of the $B^0_d$--$\overline{B^0_d}$ and 
$B^0_s$--$\overline{B^0_s}$ mixings, and the decay $K^+\to \pi^+\nu
\overline{\nu}$ \cite{AJB97}. We are optimistic that knowledge on these 
quantities will be available by the time the combined branching ratio 
$\mbox{BR}(B^\pm\to K^\pm K)$ will be measured, and look forward to 
experimental data to see how powerful the bounds on $|\Delta P|$ derived 
above are realized. In Ref.~\cite{fknp}, a different strategy to constrain
rescattering effects through the ratio $H$ has been proposed. 

Let us next have a closer look at the direct CP-violating asymmetries 
arising in the decays $B^\pm\to K^\pm K$ and $B^\pm\to\pi^\pm K$, which 
are defined by
\begin{equation}
a_{\rm CP}(B^+\to K^+\overline{K^0})\equiv
\frac{\mbox{BR}(B^+\to K^+\overline{K^0})-\mbox{BR}
(B^-\to K^-K^0)}{\mbox{BR}(B^+\to K^+\overline{K^0})+
\mbox{BR}(B^-\to K^-K^0)}
\end{equation}
\begin{equation}
a_{\rm CP}(B^+\to\pi^+ K^0)\equiv
\frac{\mbox{BR}(B^+\to\pi^+ K^0)-\mbox{BR}(B^-\to
\pi^-\overline{K^0})}{\mbox{BR}(B^+\to\pi^+ K^0)+\mbox{BR}(B^-\to
\pi^-\overline{K^0})}.
\end{equation}
Using (\ref{bspenamp}) and (\ref{bdpenamp}), as well as the $SU(3)$ flavour
symmetry as in (\ref{H-def}), it is a straightforward
exercise to derive the relation
\begin{equation}
\frac{a_{\rm CP}(B^+\to\pi^+ K^0)}{a_{\rm CP}(B^+\to K^+\overline{K^0})}=
-\,R_{SU(3)}^2\,\frac{\mbox{BR}(B^\pm\to K^\pm K)}{\mbox{BR}
(B^\pm\to\pi^\pm K)}\,\frac{(1-\lambda^2/2)\,R_b\,\sin\gamma}{R_t\,
\sin\beta}\,.
\end{equation}
Since the unitarity of the CKM matrix implies
\begin{equation}\label{UT-rel}
(1-\lambda^2/2)\,R_b\,\sin\gamma\,=\,R_t\,\sin\beta\,,
\end{equation}
where a discussion of the $(1-\lambda^2/2)$ factor can be found in \cite{BLO},
we arrive at the simple relation
\begin{equation}\label{CP-rel}
\frac{a_{\rm CP}(B^+\to\pi^+ K^0)}{a_{\rm CP}(B^+\to K^+\overline{K^0})}
\,=\,-\,R_{SU(3)}^2\,\frac{\mbox{BR}(B^\pm\to 
K^\pm K)}{\mbox{BR}(B^\pm\to\pi^\pm K)}\,,
\end{equation}
which is quite remarkable and is nicely satisfied by the results given 
in \cite{rf}.

As soon as either $a_{\rm CP}(B^+\to\pi^+ K^0)$ or $a_{\rm CP}(B^+\to 
K^+\overline{K^0})$ can be measured, we will be in a position not just to
constrain, but to determine $|\Delta P|$ and $\delta_{\Delta P}$ from $H$
and these asymmetries, if $\beta$ and $R_t$ are again used as an additional 
input. Moreover, it is possible to determine $R_{SU(3)}$ with the help of
relation (\ref{CP-rel}), providing interesting insights into $SU(3)$ 
breaking. Using (\ref{Rmin})--(\ref{rho-def}), it is even possible to
include the rescattering effects {\it completely} in the generalized bounds 
on the CKM angle $\gamma$ that are implied by the minimal value of $R$
given in (\ref{Rmin}). Interestingly, the final-state interaction effects 
may lead to a significant enhancement of the combined $B^\pm\to K^\pm K$ 
branching ratio from the ``short-distance'' value of ${\cal O}(10^{-6})$
to the $10^{-5}$ level, so that it should be possible to measure this mode
at future $B$-factories, if rescattering effects are in fact large 
\cite{defan,rf-FSI}.

A similar comment applies to the decay $B_d\to K^0\overline{K^0}$. If it 
was possible to measure the direct and mixing-induced CP asymmetries arising 
in this channel, also interesting experimental insights into the dynamics of 
penguin processes could be obtained. These CP-violating observables 
require, however, a time-dependent measurement, and can be obtained from 
the time-dependent CP asymmetry \cite{BdK0K0bar}
\begin{eqnarray}
\lefteqn{a_{\rm CP}(B_d\to K^0\overline{K^0};t)\equiv
\frac{\mbox{BR}(B^0_d(t)\to 
K^0\overline{K^0})-\mbox{BR}(\overline{B^0_d}(t)\to K^0
\overline{K^0})}{\mbox{BR}(B^0_d(t)\to K^0\overline{K^0})+
\mbox{BR}(\overline{B^0_d}(t)\to K^0\overline{K^0})}=}\nonumber\\
&&{\cal A}_{\rm CP}^{\rm dir}(B_d\to K^0\overline{K^0})\,\cos(\Delta M_d\,t)+
{\cal A}_{\rm CP}^{\rm mix-ind}(B_d\to K^0\overline{K^0})\,\sin(\Delta M_d\,t),
\end{eqnarray}
where 
\begin{equation}
{\cal A}_{\rm CP}^{\rm dir}(B_d\to K^0\overline{K^0})=\frac{1-
\left|\xi^{(d)}_{K^0\overline{K^0}}\right|^2}{1+
\left|\xi^{(d)}_{K^0\overline{K^0}}\right|^2}\,,\quad
{\cal A}_{\rm CP}^{\rm mix-ind}(B_d\to K^0\overline{K^0})=\frac{2\,
\mbox{Im}\,\xi^{(d)}_{K^0\overline{K^0}}}{1+
\left|\xi^{(d)}_{K^0\overline{K^0}}\right|^2}
\end{equation}
with
\begin{equation}
\xi^{(d)}_{K^0\overline{K^0}}=-\,e^{-i\,2\,\beta}\,
\frac{\overline{P^{(d)}}}{P^{(d)}}\,.
\end{equation}
Using the penguin amplitude (\ref{bdpenamp}), we obtain
\begin{eqnarray}
{\cal A}_{\rm CP}^{\rm dir}(B_d\to K^0\overline{K^0})&=&
\frac{2\,R_t\,|\Delta P^{(d)}|\,\sin\delta_{\Delta P}^{(d)}\,\sin\beta}{R_t^2-
2\,R_t\,|\Delta P^{(d)}|\,\cos\delta_{\Delta P}^{(d)}\,\cos\beta+
|\Delta P^{(d)}|^2}\label{CP-dirKKbar}\\
{\cal A}_{\rm CP}^{\rm mix-ind}(B_d\to K^0\overline{K^0})&=&
\frac{-\,2\,|\Delta P^{(d)}|\left[\,R_t\,\cos\delta_{\Delta P}^{(d)}-
|\Delta P^{(d)}|\,\cos\beta\,\right]\,\sin\beta}{R_t^2-
2\,R_t\,|\Delta P^{(d)}|\,\cos\delta_{\Delta P}^{(d)}
\,\cos\beta+|\Delta P^{(d)}|^2}\,.\label{CP-mixind}
\end{eqnarray}
If we look at expressions (\ref{CP-dirKKbar}) and (\ref{CP-mixind}), we 
observe that these observables depend on the three variables 
$|\Delta P^{(d)}|/R_t$, $\delta_{\Delta P}^{(d)}$ and $\beta$. Interestingly, 
the quantity $|\Delta P^{(d)}|$ parametrizing penguin topologies with internal
up and charm quarks enters in combination with $R_t$. Since $\beta$ can 
be measured through $B_d\to J/\psi\, K_{\rm S}$, the former two ``unknowns'' 
can be determined from ${\cal A}_{\rm CP}^{\rm dir}(B_d\to K^0\overline{K^0})$
and ${\cal A}_{\rm CP}^{\rm mix-ind}(B_d\to K^0\overline{K^0})$. Note
that we do not have to use any flavour symmetries to accomplish this task.

In contrast to $B^+\to K^+\overline{K^0}$, the decay $B^0_d\to K^0
\overline{K^0}$ does not receive an annihilation amplitude corresponding
to $A^{(s)}$, but contributions from ``penguin annihilation'' topologies.
Comparing ${\cal A}_{\rm CP}^{\rm dir}(B_d\to K^0\overline{K^0})$ with
$a_{\rm CP}(B^+\to K^+\overline{K^0})$, we have a probe for the importance of 
these contributions. Another important role in this respect is played by the
decay $B^0_d\to K^+K^-$, as we have already noted. If the $A^{(s)}$ amplitude
and the ``penguin annihilation'' contributions turn out to be small, a
comparison of the values for $|\Delta P^{(d)}|$ and $\delta_{\Delta P}^{(d)}$ 
obtained this way with those from the $SU(3)$ approach discussed above, 
using $H$ and direct CP violation in $B^\pm\to K^\pm K$ or 
$B^\pm\to\pi^\pm K$, we have an experimental probe for $SU(3)$ breaking. 
On the other hand, using in addition to the direct and mixing-induced 
CP asymmetries again the $SU(3)$ flavour symmetry and the observable $H$ 
specified in (\ref{H-def}), we have two options:
\begin{itemize}
\item[i)]if we use $R_t$ as an input, $\beta$ can be extracted simultaneously
with $|\Delta P|$ and $\delta_{\Delta P}$;
\item[ii)]if we use $\beta$ as an input, $R_t$ can be extracted simultaneously
with $|\Delta P|$ and $\delta_{\Delta P}$.
\end{itemize}
Following the strategies proposed in this section, it should be possible to
obtain interesting insights into the dynamics of rescattering and penguin 
processes, and in particular to take them into account in the generalized
bounds on the CKM angle $\gamma$, which are implied by $R_{\rm min}$ given
in (\ref{Rmin}). Other strategies to accomplish this task have recently been
proposed in \cite{defan,rf-FSI}, where also the uncertainties related to 
electroweak penguins \cite{groro,neubert} and their control through 
experimental data are discussed.

\boldmath
\section{A Brief Look at New Physics}\label{new-phys}
\unboldmath
Let us assume in this section that $B^0_d$--$\overline{B^0_d}$ mixing
receives significant contributions from physics beyond the Standard Model,
so that the $B^0_d$--$\overline{B^0_d}$ mixing phase is no longer given
by $2\beta$, but by a general phase (for the notation, see for instance 
\cite{HF97})
\begin{equation}
\phi_{\rm M}^{(d)}=2\beta+2\phi_{\rm new}^{(d)}\,.
\end{equation}
Since it is  unlikely that the $B_d\to J/\psi\, K_{\rm S}$ decay amplitude
is affected sizeably by new-physics contributions \cite{gl,gw}, this mode 
allows us to determine $\phi_{\rm M}^{(d)}$ from its mixing-induced CP 
asymmetry as in the case of the Standard Model analysis. 

If $B_d\to K^0\overline{K^0}$ is still governed by the Standard Model 
diagrams, its mixing-induced CP asymmetry (\ref{CP-mixind}) is modified as
\begin{eqnarray}
\lefteqn{{\cal A}_{\rm CP}^{\rm mix-ind}(B_d\to 
K^0\overline{K^0})=}\label{mixindNP}\\
&&\frac{R_t^2\,\sin
2\phi_{\rm new}^{(d)}-2\,R_t\,|\Delta P^{(d)}|\,\cos\delta_{\Delta P}^{(d)}\,
\sin\left(\beta+2\phi_{\rm new}^{(d)}\right)+|\Delta P^{(d)}|^2\sin\left
(2\beta+2\phi_{\rm new}^{(d)}\right)}{R_t^2-2\,R_t\,|\Delta P^{(d)}|\,
\cos\delta_{\Delta P}^{(d)}\,\cos\beta+|\Delta P^{(d)}|^2}\,,\nonumber
\end{eqnarray}
while the form of the direct CP asymmetry is still given by 
(\ref{CP-dirKKbar}). Note that the ``true'' value of $R_t$ enters in 
(\ref{mixindNP}). Although there are some models for new physics with a 
significant contribution to both $B^0_d$--$\overline{B^0_d}$ and 
$B^0_s$--$\overline{B^0_s}$ mixings, where their ratio still gives 
the ``true'' value of $R_t$, this need not be the case in general. In 
\cite{gnw}, a ``model-independent'' construction of the unitarity 
triangle was proposed, which holds for rather general scenarios of 
physics beyond the Standard Model. The key assumption is that new physics 
contributes significantly only to $B^0_d$--$\overline{B^0_d}$ mixing, 
as has also been done in this section. In particular, $\phi_{\rm new}^{(d)}$ 
and the ``true'' values of $R_t$ and $\beta$ can be determined this way.

An important feature of (\ref{mixindNP}) is that the
new-physics phase $\phi_{\rm new}^{(d)}$ does not enter in the combination
$\beta+\phi_{\rm new}^{(d)}$ proportional to $\phi_{\rm M}^{(d)}$, since
$\beta$ shows up also in the decay amplitude (\ref{bdpenamp}). If 
$B^\pm\to\pi^\pm K$ is also governed by the Standard Model contributions --
large CP violation in this mode, i.e.\ much larger than ${\cal O}(10\%)$, 
would indicate that this assumption does not hold -- the expression 
(\ref{H-def}) for the observable $H$ remains unchanged as well. 
Consequently, using the $SU(3)$ flavour symmetry,
$\phi_{\rm M}^{(d)}$ and the ``true'' value of $R_t$ (determined as sketched
above) as input parameters, the observables $H$, 
${\cal A}_{\rm CP}^{\rm mix-ind}(B_d\to K^0\overline{K^0})$ and direct CP 
violation in either $B_{u,d}\to K\overline{K}$ or $B^\pm\to\pi^\pm K$ allow 
the simultaneous determination of $|\Delta P|$, $\delta_{\Delta P}$, $\beta$ 
and $\phi_{\rm new}^{(d)}$. Comparing the values of $\beta$ and 
$\phi_{\rm new}^{(d)}$ obtained this way with those from 
the ``model-independent'' approach \cite{gnw}, we may find indications for
new physics if a significant disagreement should show up. If the experimental
probes for the annihilation amplitude $A^{(s)}$ and the ``penguin 
annihilation'' topologies discussed in the previous section show that these 
contributions play in fact a minor role, the most plausible interpretation 
of such a disagreement would be a new-physics contribution to 
the ``penguin-induced'' $B_d\to K^0\overline{K^0}$ decay amplitude, although 
it could in principle also originate from new-physics effects in the 
``tree-dominated'' decay $B_d\to J/\psi K_{\rm S}$. An unambiguous indication
for the latter scenario would e.g.\ be sizeable direct CP violation in 
$B_d\to J/\psi K_{\rm S}$.

\section{Conclusions}\label{concl}
In summary, we have performed an analysis of $B\to\pi K$ decays in the
presence of rescattering processes. We may distinguish between two kinds
of such final-state interaction effects. The first one is related to 
rescattering processes of the kind $B^+\to\{\overline{D^0} D_s^+\}\to\pi^+K^0$ 
and can be considered as a long-distance contribution to penguin topologies 
with internal charm quarks. Such topologies may affect the branching ratios 
for the decays $B^\pm\to\pi^\pm K$ and $B_d\to\pi^\mp K^\pm$ considerably. 
On the other hand, the second class, which is due to rescattering effects of 
the kind $B^+\to\{\pi^0K^+\}\to\pi^+K^0$, is related to penguin topologies 
with internal up quark exchanges and to annihilation processes, and does not 
affect the branching ratios significantly. However, it may lead to a sizeable 
CP-violating weak phase in the $B^+\to\pi^+K^0$ decay amplitude, and could
thereby induce CP asymmetries at the level of $10\%$ \cite{gewe}--\cite{atso}.
The corresponding rescattering effects represent an important limitation of 
the theoretical accuracy of the ``na\"\i ve'' bounds on the CKM angle 
$\gamma$ derived in \cite{fm2}. 

Moreover, we have derived isospin relations among the $B^+\to\pi^+K^0$ and
$B^0_d\to\pi^-K^+$ decay amplitudes by defining ``tree'' and ``penguin''
amplitudes in a proper way. These relations play a key role to probe $\gamma$
and allow the derivation of generalized bounds on this CKM angle. Due to a 
subtlety in implementing the $SU(2)$ isospin symmetry of strong interactions, 
the ``tree'' amplitude defined this way receives not only colour-allowed 
``tree'' contributions, but also contributions from penguin and annihilation 
topologies, and may therefore be shifted significantly from its ``factorized''
value. Interestingly, the small present central value $R=0.65$, which has 
recently been measured by the CLEO collaboration, may already indicate that 
this is actually the case. 

Instead of performing another attempt to ``calculate'' rescattering effects --
a realistic theoretical treatment is unfortunately out of reach at present -- 
we advocate to use experimental data to obtain insights into this phenomenon.
In this respect, the decays $B^\pm\to K^\pm K$ and $B^\pm\to\pi^\pm K$ are
of particular interest. Using the $SU(3)$ flavour symmetry, $\beta$ and $R_t$ 
as an input, the combined branching ratios for these modes imply a range for 
$|\Delta P|$. Measuring moreover direct CP violation in these decays -- the 
corresponding CP asymmetries can interestingly be related to the combined 
branching ratios with the help of the $SU(3)$ flavour symmetry -- both 
$|\Delta P|$ and $\delta_{\Delta P}$ can be determined. Following these
lines, the rescattering processes can be taken into account 
completely in the generalized bounds on $\gamma$ derived in \cite{defan}.
A different strategy using the $B^\pm\to K^\pm K$ and $B^\pm\to\pi^\pm K$ 
observables to accomplish this goal was proposed in \cite{defan,rf-FSI}. 

In order to obtain experimental insights into penguin processes, the decay
$B_d\to K^0\overline{K^0}$, which exhibits in contrast to the other decays
considered in this paper mixing-induced CP violation, plays also 
an important role. Combining the mixing-induced and direct CP-violating 
observables of this mode with each other, the knowledge of $\beta$ allows 
the determination of $|\Delta P^{(d)}|/R_t$ and $\delta_{\Delta P}^{(d)}$ 
without using any flavour symmetry arguments. Moreover, the importance
of annihilation and penguin annihilation topologies can be probed this
way. If these contributions should turn out to be of minor importance, 
$B_d\to K^0\overline{K^0}$ may not only shed light on the dynamics of penguin 
decays, but also on new-physics contributions to the 
$B_d\to K^0\overline{K^0}$ decay amplitude.

\vspace{0.5truecm}
\noindent
{\it Acknowledgements}

\vspace{0.3truecm}

\noindent
We are very grateful to Gerhard Buchalla, Adam Falk, Jean-Marc G\'erard,
Matthias Neubert, Yosef Nir and Daniel Wyler for stimulating discussions.

%\newpage

\boldmath
\section*{Appendix: Isospin Decomposition 
%of $B^+\to\pi^+K^0$ and $B^0_d\to\pi^-K^+$
}
\unboldmath
This appendix is devoted to the explicit derivation of the isospin 
decomposition of the $B^+\to\pi^+K^0$ and $B^0_d\to\pi^-K^+$ decay amplitudes. 
Let us to this end write the low energy effective Hamiltonian (\ref{heff}) as
\begin{equation}
{\cal H}_{{\rm eff}} = \frac{G_{\rm F}}{\sqrt{2}}\left[ 
\lambda_u^{(s)}H_u+\lambda_c^{(s)}H_c+\lambda_t^{(s)}H_{\rm P}\right],
\end{equation}
where 
\begin{eqnarray}
H_u&=&C_1(\mu) Q_1^u + C_2(\mu) Q_2^u=H_u^0+H_u^1\label{Hu-def}\\
H_c&=&C_1(\mu) Q_1^c + C_2(\mu) Q_2^c\label{Hc-def}\\
H_{\rm P}&=&-\sum^{6}_{i=3}C_i(\mu) Q_i\label{HP-def}
\end{eqnarray}
with
\begin{eqnarray} 
H_u^0&\equiv&\frac{1}{2}\left[C_1(\mu) \left(Q_1^u+Q_1^d\right) +
C_2(\mu) \left(Q_2^u+Q_2^d\right)\right]\\
H_u^1&\equiv&\frac{1}{2}\left[C_1(\mu) \left(Q_1^u-Q_1^d\right) +
C_2(\mu) \left(Q_2^u-Q_2^d\right)\right].
\end{eqnarray}
While $H_c$ and $H_{\rm P}$ correspond to $|I,I_3\rangle=|0,0\rangle$ isospin 
configurations, in the case of $H_u$ both $|0,0\rangle$ and $|1,0\rangle$
pieces are present, which are described by $H_u^0$ and $H_u^1$, respectively. 

As is well-known, the $B^+$ and $B^0_d$ mesons form an isospin doublet, i.e.\
\begin{equation}
\left|B^+\right\rangle=\left|\frac{1}{2},+\frac{1}{2}\right\rangle,\quad
\left|B^0_d\right\rangle=\left|\frac{1}{2},-\frac{1}{2}\right\rangle,
\end{equation}
whereas the isospin decomposition of the $\pi^+K^0$, $\pi^-K^+$ final states 
is given by
\begin{eqnarray}
\left|\pi^+K^0\right\rangle&=&\sqrt{\frac{1}{3}}\left|\frac{3}{2},
+\frac{1}{2}\right\rangle+\sqrt{\frac{2}{3}}\left|\frac{1}{2},
+\frac{1}{2}\right\rangle\\
\left|\pi^-K^+\right\rangle&=&\sqrt{\frac{1}{3}}\left|\frac{3}{2},
-\frac{1}{2}\right\rangle-\sqrt{\frac{2}{3}}\left|\frac{1}{2},
-\frac{1}{2}\right\rangle,
\end{eqnarray}
and contains both $I=1/2$ and $I=3/2$ components.

Taking into account that
\begin{eqnarray}
\left|1,0\right\rangle\otimes\left|\frac{1}{2},+\frac{1}{2}\right\rangle&=&
\sqrt{\frac{2}{3}}\left|\frac{3}{2},+\frac{1}{2}\right\rangle-
\sqrt{\frac{1}{3}}\left|\frac{1}{2},+\frac{1}{2}\right\rangle\\
\left|1,0\right\rangle\otimes\left|\frac{1}{2},-\frac{1}{2}\right\rangle&=&
\sqrt{\frac{2}{3}}\left|\frac{3}{2},-\frac{1}{2}\right\rangle+
\sqrt{\frac{1}{3}}\left|\frac{1}{2},-\frac{1}{2}\right\rangle
\end{eqnarray}
and using moreover the Wigner--Eckart theorem, we arrive at
\begin{eqnarray}
\left\langle K^0\pi^+\right|H_u\left|B^+\right\rangle&=&\sqrt{2}\left[
\frac{1}{3}\left({M_u^1}'-M_u^1\right)+\sqrt{\frac{1}{3}}M_u^0
\right]\label{D1}\\
\left\langle K^+\pi^-\right|H_u\left|B^0_d\right\rangle&=&\sqrt{2}\left[
\frac{1}{3}\left({M_u^1}'-M_u^1\right)-\sqrt{\frac{1}{3}}M_u^0\right],
\label{D2}
\end{eqnarray}
where
\begin{eqnarray}
M_u^0&\equiv&\left\langle\frac{1}{2},\pm\,\frac{1}{2}\right| H_u^0\left|
\frac{1}{2},\pm\,\frac{1}{2}\right\rangle\nonumber\\
M_u^1&\equiv&\mp\,\sqrt{3}\,\left\langle\frac{1}{2},\pm\,\frac{1}{2}\right| 
H_u^1\left|\frac{1}{2},\pm\,\frac{1}{2}\right\rangle\label{red-ma}\\
{M_u^1}'&\equiv&\sqrt{\frac{3}{2}}\,
\left\langle\frac{3}{2},\pm\,\frac{1}{2}\right| H_u^1\left|
\frac{1}{2},\pm\,\frac{1}{2}\right\rangle\nonumber
\end{eqnarray}
are ``reduced'' matrix elements, which carry CP-conserving strong
phases in our formalism. Since $H_c$ and $H_{\rm P}$ correspond to isospin 
singlet configurations, we have
\begin{equation}
\left\langle K^0\pi^+\right|H_c\left|B^+\right\rangle=\sqrt{\frac{2}{3}}
M_c^0,\quad \left\langle K^0\pi^+\right|H_{\rm P}\left|B^+\right\rangle=
\sqrt{\frac{2}{3}}M_{\rm P}^0
\end{equation}
\begin{equation}
\left\langle K^+\pi^-\right|H_c\left|B^0_d\right\rangle=-\sqrt{\frac{2}{3}}
M_c^0,\quad \left\langle K^+\pi^-\right|H_{\rm P}\left|B^0_d\right\rangle=
-\sqrt{\frac{2}{3}}M_{\rm P}^0,
\end{equation}
where the reduced matrix elements are defined in analogy to (\ref{red-ma}).

Combining all these equations, we arrive at
\begin{eqnarray}
\lefteqn{A(B^+\to\pi^+K^0)\equiv\left\langle K^0\pi^+\right|{\cal H}_{\rm eff}
\left|B^+\right\rangle}\nonumber\\
&&=G_{\rm F}\left[\lambda_u^{(s)}\left\{\frac{1}{3}\left(
{M_u^1}'-M_u^1\right)+\sqrt{\frac{1}{3}}M_u^0\right\}+\lambda_c^{(s)}\,
\sqrt{\frac{1}{3}}M_c^0+\lambda_t^{(s)}\,\sqrt{\frac{1}{3}}M_{\rm P}^0\right]
\equiv P^{(s)}\label{char-decom}
\end{eqnarray}
\begin{eqnarray}
\lefteqn{A(B^0_d\to\pi^-K^+)\equiv\left\langle K^+\pi^-
\right|{\cal H}_{\rm eff}
\left|B^0_d\right\rangle}\nonumber\\
&&=-\,G_{\rm F}\left[\lambda_u^{(s)}\left\{\frac{1}{3}\left(
M_u^1-{M_u^1}'\right)+\sqrt{\frac{1}{3}}M_u^0\right\}+\lambda_c^{(s)}\,
\sqrt{\frac{1}{3}}M_c^0+\lambda_t^{(s)}\,\sqrt{\frac{1}{3}}M_{\rm P}^0
\right].\label{neut-decom}
\end{eqnarray}
Comparing (\ref{char-decom}) with (\ref{ps}), it is easy to read off the
isospin decompositions of $P_{u}^{(s)}$, $P_c^{(s)}$ and $P_t^{(s)}$, while
$T^{(s)}$ has to be defined by
\begin{equation}\label{Ts-decom}
T^{(s)}\equiv \lambda_u^{(s)}\,G_{\rm F}\,\frac{2}{3}\,
\left(M_u^1-{M_u^1}'\right)
\end{equation}
in order to get the isospin relations (\ref{ampl-chariso}) and 
(\ref{ampl-neutiso}), which are at the basis of the bounds on $\gamma$ 
derived in \cite{defan}. This discussion shows nicely that these relations 
can be derived by using only isospin arguments, i.e.\ even without using the
terminology of ``penguin'' and ``tree'' contributions as is usually done
by CP practitioners. 

Let us finally note that (\ref{psu})--(\ref{pst}) and (\ref{Ts-def}), which
are expressed in terms of hadronic matrix elements of four-quark operators,
can be obtained easily from (\ref{char-decom})--(\ref{Ts-decom}) by taking 
into account that (\ref{D1}) and (\ref{D2}) yield
\begin{eqnarray}
\left\langle K^0\pi^+\right|H_u\left|B^+\right\rangle+
\left\langle K^+\pi^-\right|H_u\left|B^0_d\right\rangle&=&\frac{2}{3}\,\sqrt{2}
\left({M_u^1}'-M_u^1\right)\\
\left\langle K^0\pi^+\right|H_u\left|B^+\right\rangle-
\left\langle K^+\pi^-\right|H_u\left|B^0_d\right\rangle&=&2\,
\sqrt{\frac{2}{3}}\,M_u^0\,,
\end{eqnarray}
and using moreover (\ref{Hu-def})--(\ref{HP-def}).

\end{document}